\documentclass[11pt]{article}
\usepackage[a4paper,margin=1in]{geometry}
\usepackage{amsmath,amssymb,bm}
\usepackage{physics}
\usepackage{hyperref}
\usepackage{graphicx}
\usepackage{cite}

\title{Bose's Probabilistic Interactions, Einstein's Objections, and Their Legacy in Quantum Optics and Stochastic Mechanics}
\author{Partha Ghose \footnote{partha.ghose@gmail.com}\\Tagore Centre for Natural Sciences and Philosophy,\\ Rabindra Tirtha, New Town, Kolkata 700156, India} 
\date{}

\begin{document}
\maketitle

\begin{abstract}
In 1924, S. N. Bose proposed (i) a new counting method for photons and (ii) a probabilistic law of microscopic matter-radiation interactions, treating emission and absorption as two aspects of a single, field-dependent process. While Einstein enthusiastically extended Bose’s counting to material particles, he sharply criticized the probabilistic law, invoking detailed balance and the correspondence principle. This paper argues that (i) once one distinguishes encounter probabilities from transition rates, Einstein’s concerns can be reconciled, and (ii) that modern quantum optics and cavity QED vindicate Bose’s core intuition: ``spontaneous'' emission is not an intrinsic property of an isolated atom, as Einstein had assumed, but emerges from its coupling to the quantized field, with the rate set by the local photonic mode structure (LDOS/Purcell effect), all while satisfying Einstein’s correspondence requirement in the classical (high-intensity) limit. It is further suggested that stochastic-mechanics models--persistent random walks leading to the telegrapher’s equation, with diffusive and chiral limits yielding the Schr\"{o}dinger and Dirac equations--accord more closely with Bose's view of fundamental randomness than standard quantum mechanics, and furnish a mesoscopic bridge that reconciles micro-level stochasticity with Einstein's demand for the correct classical limit.
\end{abstract}

\section{Introduction}
In 1924 Bose wrote two papers in quick succession, one dealing with a gas of photons in which he introduced a granular phase space of the photon and a new method of counting them (Bose 1924a), and the other one dealing with photons in thermal equilibrium with matter (Bose 1924b). In the latter he first showed that Planck's law was independent of the special mechanisms of energy transfer between matter and radiation that all previous authors (Pauli (1923), Einstein and Ehrenfest (1923)) had considered till then, being determined solely by the statistics of the radiation field. He then went on to consider phenomena that defied classical expectations---the reality of atomic stationary states and the Ramsauer--Townsend reduction of low-energy electron scattering by inert gases. Interpreting these phenomena as indicating that, at quantum scales, an encounter between matter and light-quanta either leads to an exchange of energy  (an interaction) or simply ``nothing happens,'' Bose used the granular phase space he introduced in the first paper (Bose 1924a) and assigned a probability $p_r$ for an interaction when $r$ quanta are present in a phase cell. He obtained the general interaction probability
\begin{equation}
P_{\mathrm{int}}= \frac{\sum r p_r}{\sum (r + 1)p_r} =\frac{N_{\nu_r}}{A_{\nu_r}+N_{\nu_r}},
\label{eq:BosePint}
\end{equation}
with $N_{\nu_r}$ the number of quanta and $A_{\nu_r}$ the number of available cells at frequency $\nu$ and occupancy $r$. Using \eqref{eq:BosePint}, he treated atomic radiation as a single unitary process rather than two independent processes (spontaneous and induced). Writing the absorption probability as
\begin{equation}
P_{\mathrm{abs}}=\beta\frac{N_{\nu_r}}{A_{\nu_r}+N_{\nu_r}},\label{eq:Pabs}
\end{equation}
and denoting spontaneous emission strength by $\alpha$, Bose obtained the equilibrium condition
\begin{equation}
n_r \beta\frac{N_{\nu_r}}{A_{\nu_r}+N_{\nu_r}}=\alpha\,n_s \label{eq:eqbalance}
\end{equation}
which reduces to Einstein's detailed-balance (Einstein 1917) when $\alpha=\beta\,g_r/g_s$ (with $g$-factors the degeneracies). Thus, although Bose starts with only two processes (stimulated absorption and spontaneous emission), his equilibrium condition automatically generates a third process, namely `negative Einstrahlung' (Einstein's term for stimulated emission).

In the classical limit, $r \gg 1$ (and equivalently $N_{\nu_r} \gg A_{\nu_r}$),  $P_{\mathrm{int}}\to 1$, one recovers the equality of upward and downward probabilities, $\alpha n_s \simeq \beta n_r$ (oscillator-like behaviour), while in the extreme quantum limit $P_{\mathrm{int}}\to 0$ stationary states become strictly stable. For $P_{\mathrm{int}}=\varepsilon\ll 1$ the scheme naturally yields finite linewidths and random decay times. This is a situation not envisaged in Einstein's phenomenological approach.

\subsection*{Einstein's Objections and Bose's Reply}
\vspace{0.2 cm}

Einstein appended to the translated version of Bose's second paper a critical note arguing that Bose's hypothesis ran afoul of detailed balance and the correspondence principle. In thermodynamic equilibrium, he insisted, the \emph{upward} and \emph{downward} transition rates must both scale with the radiation density $N_\nu$, mirroring a classical resonator that exchanges energy symmetrically with a fluctuating field. Bose's theory has a single downward transition that does not scale with the radiation density. Hence, an external radiation cannot cause a downward transition, which violates the principle of detailed balance and the correspondence principle. He further claimed that Bose's scheme implied an unobserved $N_\nu$-dependence in the absorptivity of cold bodies. 

In a letter from Paris dated 27 January 1925 (Ghose 1994), Bose wrote:
\begin{quote}
Dear Master,

...I have been thinking of your objections all along, and so did not answer immediately. It seems to me that there is a way out of the difficulty, and I have written down my ideas in the form of a paper which I will send under a separate cover. It seems to me that the hypothesis of negative Einstrahlung stands, which as you have yourself expressed, reflects the classical behaviour of a resonator in a fluctuating field. But the additional hypothesis of a spontaneous change, independent of the state of the field, seems to me not necessary. I have tried to look at the radiation field from a `new' standpoint, and have sought to separate the propagation of the Quantum of energy from the propagation of the electromagnetic influence. ...
\end{quote}

What could Bose have meant by ``the hypothesis of negative Einstrahlung stands''? The answer possibly lies in the distinction between probabilities and rates.

\subsection*{From probabilities to \emph{rates}: a minimal kinetic closure}

Eq.~\eqref{eq:BosePint} is a \emph{conditional} probability per encounter; it does not fix how often encounters occur.
A natural kinetic identification—already hinted by Bose's counting—is to take the \emph{encounter frequency} proportional to the total number of ``opportunities'' for interaction,
i.e.\ to the \emph{denominator} in \eqref{eq:BosePint}:
\begin{equation}
\Gamma_\nu \;=\; \kappa\,\mathcal{O}_\nu,
\qquad
\mathcal{O}_\nu := A_{\nu_r}+N_{\nu _r},
\label{eq:Gamma}
\end{equation}
with a kinetic constant $\kappa$ setting the time scale.
Then the total interaction \emph{rate} per molecule is
\begin{equation}
R_{\mathrm{int}} \;=\; \Gamma_\nu\,P_{\mathrm{int}}
\;=\; \kappa\,(A_{\nu_r}+N_{\nu_r})\,\frac{N_{\nu_r}}{A_{\nu_r}+N_{\nu r}}
\;=\; \kappa\,N_{\nu_r}.
\label{eq:Rint}
\end{equation}
Thus, \emph{independently of intensity}, the overall interaction rate is \emph{linear} in $N_{\nu_r}$.

Decomposing \eqref{eq:Rint} into absorption and (stimulated) emission via \eqref{eq:Pabs} (and its emission analogue with the same $\beta$ in Bose’s scheme) gives the \emph{per-molecule} rates
\begin{align}
R_{\mathrm{abs}} &= \Gamma_\nu\,P_{\mathrm{abs}} \;=\; \kappa\,\beta\,N_{\nu_r}, \label{eq:Rabs}\\
R_{\mathrm{em,stim}} &= \Gamma_\nu\,P_{\mathrm{em,stim}} \;=\; \kappa\,\beta\,N_{\nu_r}, \label{eq:Remstim}
\end{align}
while the spontaneous piece is $R_{\mathrm{sp}}=\alpha$ (independent of $N_{\nu_r}$).
On a population $n_r$ (resp.\ $n_s$), the upward (resp.\ downward) \emph{ensemble} rates are $n_r R_{\mathrm{abs}}$ and $n_s(R_{\mathrm{em,stim}}+R_{\mathrm{sp}})$, and the equilibrium condition \eqref{eq:eqbalance} follows as in Bose’s derivation.

Einstein demanded that, in the classical limit, both upward \emph{and} downward \emph{rates} scale with radiation density $N_\nu$ (as for a classical resonator in a fluctuating field).
Equations \eqref{eq:Rabs}–\eqref{eq:Remstim} satisfy this requirement \emph{identically}:
for any intensity, and in particular as $N_{\nu r}\gg A_{\nu r}$ (classical regime), one has
\[
R_{\mathrm{abs}}\propto N_{\nu r},\qquad
R_{\mathrm{em,stim}}\propto N_{\nu r},\qquad
R_{\mathrm{sp}}\;\text{subleading for large }N_{\nu r}.
\]
Hence Bose’s probability law \eqref{eq:BosePint}, together with the minimal kinetic closure \eqref{eq:Gamma} already suggested by his counting, \emph{recovers the correspondence limit for rates}.

\subsection*{Cold-body absorptivity}

Einstein also worried that Bose’s law would imply an anomalous dependence of a cold body’s absorptivity on $N_\nu$ at low intensity.
But \eqref{eq:Rabs} shows that the absorption rate is \emph{linear} in $N_{\nu_r}$ even for $N_{\nu_r}\ll A_{\nu_r}$; there is no quadratic low-intensity anomaly once the encounter frequency is tied to the same opportunity count that normalizes $P_{\mathrm{int}}$.
In this kinetic reading, Bose’s scheme matches the standard linear-response phenomenology.

Hence, Bose’s Eq.~\eqref{eq:BosePint} already enforces classical symmetry \emph{per encounter} ($P_{\mathrm{int}}\to 1$ as $r\to\infty$).
Making explicit the encounter frequency as in \eqref{eq:Gamma} converts probabilities into rates that are linear in $N_\nu$, thereby (i) reproducing Einstein’s correspondence limit and (ii) avoiding the cold-body absorptivity concern. In this sense Bose’s proposal does not call for abandoning Einstein’s hypothesis of negative Einstrahlung (stimulated emission)--it stands.

\subsection*{Where the disagreement lay (textual evidence)}

The kinetic reconciliation proposed above (probabilities $\to$ rates) makes it unlikely that the dispute was about algebra; rather, it reflected, in my opinion, priors Einstein held in the 1917–1926 period:
\begin{enumerate}

\item \emph{Three elemental transition processes}: In his 1917 paper Einstein postulates \emph{three} elemental processes: spontaneous emission with probability per unit time $A_{mn}\,dt$ (independent of radiation density), and stimulated absorption/stimulated emission with probabilities $B_{mn}\rho(\nu)\,dt$ and $B_{nm}\rho(\nu)\,dt$ that scale with the spectral energy density $\rho(\nu)$ \cite{Einstein1917}. As he puts it in the English translation, spontaneous emission occurs ``\emph{independent of whether it is excited by an external field or not}.'' This makes spontaneity an \emph{intrinsic} property of the atom in Einstein’s kinetics.

\item \emph{Aversion to fundamental randomness}: In his correspondence with Born (4 December 1926), Einstein wrote: ``\emph{I am at all events convinced that He does not play dice'' (Born-Einstein Letters 1971)}, making explicit his resistance to a theory grounded in irreducible chance.
\end{enumerate}

Given these priors, Bose's hypothesis--two elemental processes only, and intrinsically probabilistic interactions--was unacceptable to Einstein.
 
\section{Spontaneity as Emergent in Quantum Optics}
Modern Quantum Optics (Scully and Zubairy 1997) resolves the impasse in a way that largely vindicates Bose's 1925 insight, namely that spontaneous emission is field-dependent. For a two-level emitter coupled to the quantized field, Fermi's Golden Rule yields
\begin{equation}
W_{\mathrm{em}}\propto \big(N_\omega+1\big)\,|g_\omega|^2,\qquad
W_{\mathrm{abs}}\propto N_\omega\,|g_\omega|^2,
\end{equation}
where the celebrated ``$+1$'' is the vacuum term originating in the field commutator $[a_{\mathbf{k}},a_{\mathbf{k}}^\dagger]=1$. In the classical limit $N_\omega\gg 1$ (e.g.\ a strong coherent state) one has $N_\omega+1\simeq N_\omega$, so both up/down rates scale with intensity as Einstein required. A compact Markovian formulation makes the crossover explicit:
\begin{equation}
\dot{\rho}=-\frac{i}{\hbar}[H,\rho]
+\gamma\big(N(\omega_0)+1\big)\,\mathcal{D}[\sigma_-]\rho
+\gamma N(\omega_0)\,\mathcal{D}[\sigma_+]\rho,
\quad
\mathcal{D}[L]\rho=L\rho L^\dagger-\tfrac{1}{2}\{L^\dagger L,\rho\}.
\end{equation}
with $\mathcal{D}[L]\rho=L\rho L^\dagger-\tfrac12\{L^\dagger L,\rho\}$. Vacuum ($N=0$) leaves the $+1$ term---``spontaneous emission''---as a quantum residue of the same coupling that drives stimulated processes; large $N$ recovers the symmetric classical exchange envisaged by Einstein.

\subsection*{Cavity QED: LDOS and Purcell Enhancement/Inhibition}
Cavity QED (Cohen-Tannoudji (1992), Gardiner and Zoller (2004), Haroche (2006), Kleppner (1981), Novotny and Hecht (2012), Purcell (1946),  Yablonovitch (1987)) reveals that the ``spontaneous'' rate is not an intrinsic atomic constant but a property of the emitter \emph{in} its electromagnetic environment. The rate is proportional to the local density of electromagnetic states (LDOS),
\begin{equation}
\Gamma(\mathbf{r},\omega_0)\propto \mathbf{d}\cdot \mathrm{Im}\,\mathbf{G}(\mathbf{r},\mathbf{r};\omega_0)\cdot \mathbf{d}\;\propto\;\rho(\mathbf{r},\omega_0),
\end{equation}
with $\mathbf{G}$ the dyadic Green function. Engineering the mode structure (high-$Q$ low-$V$ cavities, photonic band gaps) enhances or inhibits emission---the Purcell effect,
\begin{equation}
F_{\mathrm{P}}=\frac{3}{4\pi^2}\left(\frac{\lambda}{n}\right)^{\!3}\frac{Q}{V}\,\xi(\mathbf{r},\hat{\mathbf{d}}),
\end{equation}
where $\xi$ accounts for spatial/polarization overlap. This is a concrete realization of Bose's ``cells'' picture: change the number/weight of accessible modes, and you change the interaction probability and the emission rate.

\section{Stochastic Mechanics as a Bridge}
\label{sec:SM-bridge}

A natural kinematic realization of Bose’s idea that at the microlevel an encounter may
result in an interaction or in “nothing happening” is a \emph{persistent random motion}:
ballistic flight segments interspersed with rare, memoryless (Poisson) reorientations.
This injects microscopic randomness without abandoning continuity or finite signal speed,
and—upon coarse graining—yields familiar transport and wave equations.

Consider a particle on $\mathbb{R}$ moving at fixed speed $c$ and reversing direction as a
Poisson process of rate $\lambda>0$ (Kac (1974)). Let $P_\pm(x,t)$ be the right/left moving
probability densities. The forward equations (\emph{Kac system}) are
\begin{align}
\partial_t P_+ &= -\,c\,\partial_x P_+ \;-\; \lambda P_+ \;+\; \lambda P_-,
\label{eq:kac+}\\
\partial_t P_- &= \;\;\,c\,\partial_x P_- \;-\; \lambda P_- \;+\; \lambda P_+.
\label{eq:kac-}
\end{align}
Define $\rho:=P_++P_-$ and $j:=c(P_+-P_-)$. Adding/subtracting
\eqref{eq:kac+}–\eqref{eq:kac-} gives
\begin{align}
\partial_t \rho + \partial_x j &= 0,
\label{eq:cont}\\
\partial_t j + 2\lambda\, j &= c^2\,\partial_x \rho.
\label{eq:mom}
\end{align}
Eliminating $j$ yields the (hyperbolic) \emph{telegrapher’s equation} (TE):
\begin{equation}
\partial_{tt}\rho \;+\; 2\lambda\,\partial_t \rho \;=\; c^2\,\partial_{xx}\rho.
\label{eq:TE}
\end{equation}
Two regimes are immediate: (i) \emph{ballistic} for $t\!\ll\!\lambda^{-1}$ and
(ii) \emph{diffusive} for $t\!\gg\!\lambda^{-1}$, where \eqref{eq:TE} reduces to
\begin{equation}
\partial_t \rho \;=\; D\,\partial_{xx}\rho,
\qquad
D \;=\; \frac{c^2}{2\lambda}.
\label{eq:diffusion}
\end{equation}

\subsection*{Diffusive limit and Nelson’s route to Schr\"odinger's equation}
\vspace{0.2 cm}

Nelson’s stochastic mechanics (Nelson (1966)) identifies $D=\hbar/(2m)$ and models the particle
trajectory $X_t$ by forward/backward It\^o diffusions with drifts $b_\pm$.
Let $\rho$ be the density of $X_t$ and define the \emph{current} and \emph{osmotic}
velocities by
\begin{equation}
v := \frac{b_+ + b_-}{2},
\qquad
u := \frac{b_+ - b_-}{2} \;=\; D\,\nabla \ln \rho.
\label{eq:vu}
\end{equation}
Then
\begin{align}
\partial_t \rho + \nabla\!\cdot(\rho v) &= 0,
\label{eq:continuity}\\
\partial_t S + \frac{(\nabla S)^2}{2m} + V
\;-\; \frac{\hbar^2}{2m}\frac{\nabla^2\sqrt{\rho}}{\sqrt{\rho}} &= 0,
\label{eq:stochHJ}
\end{align}
for a real phase $S$ (defined up to a constant). The first of this pair of equations is a continuity equation and the second a Hamilton-Jacobi equation with an additional stochastic term, the quantum potential. These are coupled differential equations which can be combined into a single equation for a complex wave function $\psi=\sqrt{\rho}\,e^{iS/\hbar}$ that satisfies the Schr\"odinger equation
\begin{equation}
i\hbar\,\partial_t \psi \;=\; -\frac{\hbar^2}{2m}\,\nabla^2\psi + V\psi.
\label{eq:Sch}
\end{equation}
Thus, a \emph{random} microdynamics (Poisson reversals $\to$ diffusion) yields the
\emph{unitary} wave equation while automatically preserving equilibrium and detailed balance
at the ensemble level.

\subsection*{Chiral persistent walks and the Dirac equation in \(1{+}1\)}
\vspace{0.2 cm}

Let $\psi_R,\psi_L$ denote right/left \emph{amplitudes}. A chiral persistent walk with
Poisson chirality flips (with rate $\lambda$) and coherent mass coupling
($mc^2/\hbar$) reads
\begin{align}
(\partial_t + c\,\partial_x)\psi_R &= -\lambda\,\psi_R + \lambda\,\psi_L
\;-\; i\,\frac{mc^2}{\hbar}\,\psi_R,\\
(\partial_t - c\,\partial_x)\psi_L &= -\lambda\,\psi_L + \lambda\,\psi_R
\;-\; i\,\frac{mc^2}{\hbar}\,\psi_L.
\end{align}
In the coherent limit this is equivalent to the Dirac equation in \(1{+}1\) (Gaveau (1984))  for
$\Psi=(\psi_R,\psi_L)^{\!\top}$,
\begin{equation}
i\hbar\,\partial_t \Psi \;=\; \big(-i\hbar c\,\alpha\,\partial_x + \beta\,mc^2\big)\Psi,
\qquad \alpha=\sigma_x,\;\beta=\sigma_z.
\label{eq:Dirac1D}
\end{equation}
Elimination of one chirality reproduces a telegrapher's equation (TE) with a mass term.

\subsection*{Emergent linewidths and waiting times from rare events}

Rare stochastic events with rate $\varepsilon\ll 1$ lead to exponential waiting-time
distributions $p(\tau)=\varepsilon e^{-\varepsilon \tau}$ and Lorentzian lineshapes with width
$\sim\varepsilon$ in frequency space—providing a transparent mechanism for natural linewidths
and random decay times without \emph{ad hoc} spontaneity. In master-equation form they
appear as weak Lindblad terms coexisting with Hamiltonian evolution.

\subsection*{Classical correspondence from hydrodynamic limits}

Einstein’s correspondence requirement is naturally met: hydrodynamic/long-time limits of
the stochastic dynamics yield classical transport (diffusion/drift–diffusion), while in
\eqref{eq:stochHJ} the quantum potential vanishes as $\hbar\to 0$ (i.e.\ $D\to 0$),
reducing to classical Hamilton–Jacobi plus Liouville continuity (classical statistical mechanics). In the chiral case, weak
mixing and large occupations recover the classical oscillator/wave picture.

\subsection*{Measurement and collapse}

In Nelson-type stochastic mechanics collapse is \emph{predicted}, not postulated as in standard quantum mechanics (Pavon (1999)).
The state update upon measurement is obtained by
\emph{conditioning} the diffusion on the registered outcome and selecting—via a stochastic
variational/Schr\"odinger-bridge (Doob $h$-transform)—the closest process to the pre-measurement
dynamics that satisfies the constraint. As shown by Pavon, this yields the
von Neumann–L\"uders rule
$\psi \mapsto P\psi/\|P\psi\|$ without an independent “collapse postulate,”
defusing the usual measurement paradoxes (Cat, Wigner’s Friend) tied to an ad hoc
projection.

\subsection*{Relativistic extension to \(3{+}1\) Minkowski space}

The random-flight $\!\to$ TE $\!\to$ wave-equation bridge extends to full spacetime.
Gaveau et al (Gaveau (1984)) constructed a Poissonian kinematics whose real form yields the
telegrapher-type equation with finite signal speed and whose analytic continuation
produces the Dirac equation, accommodating \(3{+}1\) Minkowski structure.

Thus, a single microscopic ingredient—\emph{probabilistic kinematics with rare events}—yields the telegrapher's equation
(TE) and, via Nelson’s identification $D = \hbar/2m$, the full Schr\"{o}dinger dynamics.”. A chiral version yields the Dirac equation in the chiral (Weyl) representation. This provides a concrete, mechanistic bridge from Bose’s probabilistic intuition to unitary wave equations while preserving classical correspondence in the appropriate limits.

\section*{Acknowledgements}
Historical details and portions of Sections 1–2 draw on my earlier article (Ghose 1994) and on materials compiled in \emph{S. N. Bose: The Man and His Work, Part I (S. N. Bose National Centre for Basic Sciences)}. I am deeply grateful to Professor S. N. Bose for many personal conversations over a long period (1964--1973) in which he patiently clarified the subtleties of his disagreement with Einstein and urged me to seek a resolution in the light of modern developments. Though no one can tell now how he would have received this resolution, I offer it as my best homage during the centennial of the birth of quantum mechanics. Any weaknesses or oversights are mine alone.

\bibliographystyle{unsrt}

\begin{thebibliography}{99}

\bibitem{BBirula}
 Bia{\l}ynicki-Birula, I. (1996), ``Photon Wave Function,'' \emph{Progress in Optics}, Vol. XXXVI, Ed. E. Wolf, (Elsevier, Amsterdam).

\bibitem{Bose1924a}
Bose, S.\,N.\ (1924a) ``Plancks Gesetz und Lichtquantenhypothese,'' \emph{Zeitschrift f\"ur Physik} \textbf{26}, 168--171.

\bibitem{Bose1924b}
Bose, S.\,N.\ (1924b), ``W\"armegleichgewicht im Strahlungsfeld bei Anwesenheit von Materie,'' \emph{Zeitschrift f\"ur Physik} \textbf{27}, 384--393.

\bibitem{BornEinsteinLetters}
Born, M. and Einstein, A. (1971) \emph{The Born--Einstein Letters: Correspondence Between Albert Einstein and Max and Hedwig Born from 1916-1955, with Commentaries by Max Born}, Letter 27 
transl.\ I.~Born (Macmillan, London).

\bibitem{CohenTannoudji}
Cohen-Tannoudji, C., Dupont-Roc, J. and Grynberg, G. (1992), \emph{Atom--Photon Interactions: Basic Processes and Applications} (Wiley).

\bibitem{Einstein1917}
Einstein, A. (1917), ``Zur Quantentheorie der Strahlung,'' \emph{Physikalische Zeitschrift} \textbf{18}, 121--128.

\bibitem{Einstein1923}
Einstein, A. and Ehrenfest, P. (1923), \emph{Zeitschrift f\"ur Physik} \textbf{19}, 301.

\bibitem{Gaveau1984}
Gaveau, B., Jacobson, T., Kac, M. and Schulman, L.\ S. (1984), ``Relativistic Extension of the Analogy between Quantum Mechanics and Brownian Motion,'' \emph{Phys.\ Rev.\ Lett.} \textbf{53}, 419--422.

\bibitem{GardinerZoller}
Gardiner, C.\ W. and Zoller, P. (2004), \emph{Quantum Noise}, 3rd ed.\ (Springer).

\bibitem{BoseCollected}
Ghose, P. (1994), ``Bose Statistics: a historical perspective'' in \emph{S. N. Bose: The Man and His Work, Part I: Collected Scientific Papers}, pp 35--71 (sections n and o, pp 57-68), https://www.bose.res.in/Prof.S.N.Bose-Archive/


\bibitem{HarocheRaimond}
Haroche, S. and Raimond, J.-M. (2006), \emph{Exploring the Quantum: Atoms, Cavities, and Photons} (Oxford University Press).

\bibitem{KacTelegraph}
Kac, M. (1974), ``A Stochastic Model Related to the Telegrapher's Equation,'' \emph{Rocky Mountain J.\ Math.} \textbf{4}, 497--509.

\bibitem{Kleppner}
Kleppner, D. (1981), ``Inhibited Spontaneous Emission,'' \emph{Phys.\ Rev.\ Lett.} \textbf{47}, 233--236.

\bibitem{Nelson1966}
Nelson, E. (1966), ``Derivation of the Schr\"odinger Equation from Newtonian Mechanics,'' \emph{Phys.\ Rev.} \textbf{150}, 1079--1085.

\bibitem{NovotnyHecht}
Novotny, L. and Hecht, B. (2012), \emph{Principles of Nano-Optics}, 2nd ed.\ (Cambridge University Press), chs.\ 8--9.

\bibitem{Pauli1923}
Pauli, W. (1923), \emph{Zeitschrift f\"ur Physik} \textbf{18}, 272.

\bibitem{Pavon1999}
Pavon, M. (1999), ``Stochastic mechanics and the Feynman integral,''
\emph{J.\ Math.\ Phys.} \textbf{40}, 5565--5577.

\bibitem{Purcell}
Purcell, E.\ M. (1946), ``Spontaneous Emission Probabilities at Radio Frequencies,'' \emph{Phys.\ Rev.} \textbf{69}, 681.

\bibitem{ScullyZubairy}
Scully, M.\ O. and Zubairy, M.\ S. (1997), \emph{Quantum Optics} (Cambridge University Press).

\bibitem{Yablonovitch}
Yablonovitch, E. (1987), ``Inhibited Spontaneous Emission in Solid-State Physics and Electronics,'' \emph{Phys.\ Rev.\ Lett.} \textbf{58}, 2059--2062.

\end{thebibliography}

\end{document}